\documentclass[aps,twocolumn,superscriptaddress,prl]{revtex4}

\usepackage{amssymb}
\usepackage{amsmath}
\usepackage{epsfig}
\usepackage[english]{babel}
\usepackage{hyperref}

\emergencystretch=\maxdimen
\begin{document}

\title{Pulsed interaction signals as a route to biological pattern formation}

\author{Eduardo H.  Colombo}
\email{ecolombo@princeton.edu}
\affiliation{Department of Ecology \& Evolutionary Biology, Princeton University, Princeton, NJ 08544, USA\looseness=-1}
\affiliation{Department of Ecology, Evolution, and Natural Resources, Rutgers University, New Brunswick, NJ 08901, USA\looseness=-1}
\affiliation{Instituto de F\'{\i}sica Interdisciplinar y
Sistemas Complejos (IFISC), CSIC-UIB, Campus Universitat Illes
Balears, 07122, Palma de Mallorca, Spain }

\author{Crist\'obal
L\'opez}
\email{clopez@ifisc.uib-csic.es}

\author{Emilio
Hern\'andez-Garc\'{\i}a} \email{emilio@ifisc.uib-csic.es}
\affiliation{Instituto de F\'{\i}sica Interdisciplinar y
Sistemas Complejos (IFISC), CSIC-UIB, Campus Universitat Illes
Balears, 07122, Palma de Mallorca, Spain }

\begin{abstract}
We identify a mechanism for biological spatial pattern
formation arising when the signals that mediate interactions
between individuals in a population have pulsed character. Our
general population-signal framework shows that while for a slow
signal-dynamics limit no pattern formation is observed for any
values of the model parameters, for a fast limit, on the
contrary, pattern formation can occur. Furthermore, at these
limits, our framework reduces, respectively, to
reaction-diffusion and spatially nonlocal models, thus bridging
these approaches.
\end{abstract}

\date{\today}

\maketitle

\textit{Introduction.---} One of the striking manifestations of
self-organization in complex systems is the emergence of
regular spatial patterns at scales much larger than the ones
associated to the individual
components~\cite{rietkerk2008regular}. In biological
populations this phenomenon has been observed in many contexts
including semi-arid
vegetation~\cite{Klausmeier1999,von2001,Fernandez-Oto2014},
bird swarms~\cite{vicsek,attanasi2014information} or bacteria
colonies~\cite{ben1990,tyson1999minimal}. Besides being
fascinating, pattern formation has been shown to critically
affect the stability and resilience of
ecosystems~\cite{rietkerk_self-organized_2004,bonachela_termite_2015}.

Behind the mechanisms
responsible for self-organization there is often
 an agent or substance  working as a signal
 that mediates the interactions.
Signals have distinct emission protocols, propagation dynamics
and occur in a wide range of temporal and spatial
scales~\cite{smith2003animal}. For example, species might use
acoustic~\cite{ricardo2013p}, visual~\cite{caro2017} or
chemical~\cite{ojalvo2018} signals to attract, repel,  harm or
support targeted individuals. It is this exchange of signals
and the details of its dynamics that ultimately drive
self-organization
process~\cite{giuggioli2011animal,potts2016memory} and,
consequently, control other key macroscopic
outcomes~\cite{niehaus2019microbial,rietkerk_self-organized_2004}.

Despite the numerous studies analyzing how interactions control
pattern formation, the focus has been mostly on continuous and
smooth signal dynamics. This overlooks interactions that are
mediated by flashing pulsed signals. Therefore, how this
fine-scale dynamics scales-up affecting pattern formation is
poorly understood. Here we show that a timescale transition
from slow (smooth) to fast (pulsed) signal dynamics creates a
route to pattern formation alternative to the most studied ones
arising from Turing-like mechanisms~\cite{Turing1952a}.

This finding is obtained by studying a general
activator-inhibitor (population-signal) model, where a
population interacts through the release of  harmful signals.
Our study extends standard activator-inhibitor
structure~\cite{gierer1972,murray2003}, by explicitly
describing the fine-scale dynamics associated with the release
and spreading of signals. This framework recovers two distinct
structures at the regimes of slow and fast signal dynamics that
can lead to qualitative changes in spatial stability. For slow signal  (with timescales similar to those of
the population), we recover a standard reaction-diffusion
system which, for a broad set of population and
signal dynamics, does not exhibit Turing instability
for any values of model parameters. For the same
system dynamics,  but with sufficiently
fast signals, the system can be described by a single
integrodifferential equation, where the toxic effects are
captured by a competitive nonlocal spatial interaction. In this
limit, spatial instability can occur leading to pattern
formation.

Since we explicitly derive the underlying interference
competition mechanism behind the nonlocal effective
description, these results address a long-standing shortcoming:
that paradigmatic nonlocal models of competitive type leading
to spatial patterns have been usually proposed
phenomenologically with no systematic
derivation~\cite{Fuentes2003,Borgogno2009,Fernandez-Oto2014}.
In the cases in which such derivation has been provided, the
resulting equation did not have the characteristics needed for
pattern-forming instabilities (see, for example
Ref.~\cite{ricardo2014})

\textit{Model.---} Our aim is to model an ensemble of simple
organisms in a one-dimensional spatial domain (we do not expect
this dimensional restriction to be essential for our results).
They move, reproduce and release harmful signals in the form of
pulses. These pulses can have biochemical origins, such as a
toxic substance, but can also be physical, in the form of
electricity, heat, sound and light, which can compromise the
targets' survival, and lead to a competing dynamics among the
individuals~\cite{foster2013,burt1943,caro2017}. We describe
this scenario at the population-level by the following general
density-field description,
\begin{align}
\label{geneq1}
\tau_\rho\partial_t \rho &= \mathcal{L}( \rho,\partial_x\rho)  - \epsilon \rho\phi\, ,\\
\tau_\phi\partial_t \phi &= L( \phi,\partial_x\phi) + \mathcal{R}_\rho(x,t)\, ,
\label{geneq2}
\end{align}
where $\rho$ and $\phi$ are the population density and signal
intensity, respectively. $\mathcal{L}$ and $L$ give the
population and signal dynamics when uncoupled, including
diffusion or other transport processes; $\tau_\rho$ and
$\tau_\phi$ explicitly set the timescales for the population
and signal dynamics, respectively; and $\epsilon$ is an
exposition factor related to the population sensitivity to the
toxin, which is released according to $\mathcal{R}_\rho$.

We consider that signal releases occur in pulses that are
controlled by the population density in the following manner:
Their starting time-space locations $\{t_i,x_i\}$ are
independent Poisson random events with a probability of
occurring within small intervals $dx$ and $dt$ given by
$\alpha\rho(x,t) dx dt$. The pulses have duration
${\bar\delta}$, negligible spatial extent, and equal
intensities $I_0$:
\begin{align}
\label{releases}
\mathcal{R}_\rho(x,t) &= \sum_i I_0 \Pi_{\bar\delta}(t-t_i)\delta(x-x_i)\, ,
\end{align}
where $\Pi_{\bar\delta}(t)$ is the indicator function of the
time interval $[0,\bar\delta]$. The expected inter-event time,
$\langle t_{i+1}-t_i\rangle$, is given by $\tau_R =
1/(N(t)\alpha)$, where $N(t)=\int_{-\infty}^{+\infty}\rho(x)
dx$ is the total population size.
Eqs.~(\ref{geneq1}-\ref{releases}) together establish the model
studied in this work, being constituted by a continuous
population model but with a pulsed spatiotemporal dynamics for
the signal~\cite{grimaCurrent}.

The characteristic timescales are, besides $\tau_\rho$ and
$\tau_\phi$, the duration of the pulses, ${\bar\delta}$, and
the mean pulse inter-event time, $\tau_R$. We will focus on
cases in which pulse duration is much shorter than release
inter-event time, which is itself much shorter than population
dispersal and other demographic processes, $\bar{\delta} \ll
\tau_R \ll \tau_\rho$. This means that there is a timescale
separation between interaction events and their consequences to
population dynamics.

\begin{figure}[h]
\includegraphics[width=\columnwidth]{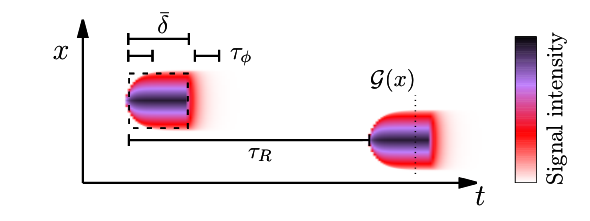}
\caption{Schematic representation of the timescales of the signal $\phi(x,t)$ in
the fast signal-dynamics limit ($\tau_\phi \to 0$). Signal response
to two signal release pulses is shown for moderately fast signals. $\phi$ remains mainly
localized in time within the pulse duration $\bar\delta$ (dashed rectangle). The rising and decaying
parts of the signal are indicated by the two short
segments close to the label $\tau_\phi$. The vertical dotted line indicates an
intermediate time at which
signal intensity attains the steady $\mathcal{G}$ profile.
In the plot, the lapse between pulses is set to
$\tau_R$ (the mean inter-event time) and the field $\phi$ spreads according to Eq.~(\ref{choice})
(with $\nu=4$ and $\mu=1$ as in Fig.~\ref{fig:poisson_patterns}b).
}
\label{fig:scheme}
\end{figure}

In the following, we investigate how the system spatial
stability changes as a function of the signal timescale,
$\tau_\phi$. We obtain effective descriptions for the
population-toxin dynamics and the respective pattern forming
conditions for a) the \emph{slow signal-dynamics limit}, in
which the toxin-field relaxation is slow, being comparable to
population dynamics timescales $\tau_\phi/ \tau_\rho \sim 1$,
and thus $\bar{\delta}, \tau_R \ll \tau_\phi$; and b)
\emph{fast signal-dynamics limit}, when signal response is the
faster of all the timescales, $\tau_\phi \ll
\bar{\delta},\tau_R,\tau_\rho$ (see Fig.~\ref{fig:scheme}).

\emph{Slow signal-dynamics limit.---} When $\tau_\phi/\tau_\rho
\sim 1$ the inter-event release time is much shorter than
population and signal timescales, $\tau_R \ll \tau_\phi,
\tau_\rho$. Then, the toxin field $\phi$ in Eq.~(\ref{geneq2})
feels the average of the toxin release pulses, which are many
and occur too fast for $\phi$ to follow them. Consequently, we
can replace $\mathcal{R}_\rho$ by its average over small time
windows $\Delta t \ll \tau_\phi$ and small vicinities $\Delta
x$:
\begin{align}
\label{avgdef}
\langle \mathcal{R}_\rho(x,t) \rangle \equiv
\frac{1}{\Delta t \Delta x} \int_{-\Delta t}^{0}\int_{-\frac{\Delta x}{2}}^{\frac{\Delta x}{2}}\mathcal{R}_\rho(x+x',t+t') dx' dt' .
\end{align}

Using Eq. (\ref{releases}), $\langle \mathcal{R}_\rho(x,t)
\rangle = (\Delta x\Delta t)^{-1}\sum_{i=1}^{n_R} I_0
\bar\delta  =  I_0 \bar\delta n_R/(\Delta x \Delta t)$, where
$n_R$ is the number of pulses that have occurred during the
considered space-time window. Noting that pulses are
independent events, $n_R$ for each cell follows a Poisson
distribution with mean $\alpha\rho(x,t)\Delta x \Delta t$. If
$\Delta t$ is chosen sufficiently large (but still much smaller
than $\tau_\phi$) $n_R$ becomes large and its coefficient of
variation (ratio of standard deviation to mean) vanishes so
that fluctuations can be neglected. Thus $n_R\approx
\alpha\rho(x,t)\Delta x \Delta t$. As a consequence, in the
slow signal limit, $\langle \mathcal{R}_\rho(x,t)\rangle
\approx I_0 {\bar\delta} \alpha \rho(x,t)$. 
The other terms in Eqs. (\ref{geneq1}-\ref{geneq2}) can also be
coarse-grained but, due to their slow response times, they
remain constant and unaffected by the procedure: $\langle \rho
\rangle \simeq \rho$, $\langle \phi \rangle \simeq \phi$,
$\langle \phi\rho  \rangle \simeq \phi\rho$.

\emph{Fast signal-dynamics limit.---} In this fast limit,
$\tau_\phi/\tau_\rho \to 0$, the signal dynamics is much faster
than any other process. Then, we can expect the signal field to be
always in constant equilibrium with the release events: it
immediately reaches a fixed stationary profile,
$\mathcal{G}(x)$ during the pulse duration,
$0<t-t_i<{\bar\delta}$, and dissipates immediately when release
ceases (we assume a $L$ dynamics that leads to signal
dissipation in the absence of releases). In Fig.
\ref{fig:scheme}, we present a schematic representation of the
fast signal propagation, highlighting with a dashed rectangle
the area in which signals would be confined taking
$\tau_\phi\to 0$. The profile $\mathcal G$ at intermediate
times (such as the vertical dotted line in Fig.
\ref{fig:scheme}) can be obtained by solving Eq. (\ref{geneq2})
under the limit $\tau_\phi\rightarrow 0$. For a single pulse in
(\ref{releases}) at $x=0$,
$L(\mathcal{G},\partial_x\mathcal{G})+ I_0\delta(x)=0$. The
conditions $\tau_\phi \ll \bar\delta \ll \tau_R$ guarantee that
pulses are non-overlapping, so that the solution of Eq.
(\ref{geneq2}) can be built just adding up the successive
responses to the different pulses: $\phi(x,t) \approx \sum_i
\mathcal{G}(x-x_i)\Pi_{\bar\delta}(t-t_i)$. We now perform, as
in Eq. (\ref{avgdef}), an average of Eq. (\ref{geneq1}) over
small intervals $\Delta x$ and $\Delta t \ll \tau_\rho$.
Because of timescale separation, all terms remain unaltered
except the last one containing $\phi$, which becomes
$\left<\rho \phi\right>\approx \rho \left<\phi\right>$.
Calculation of this last average is performed in detail in the
Supplemental Material \cite{SM}, with the final result
$\left<\phi\right>\approx {\bar\delta}\alpha
[\mathcal{G}\ast\rho]$, where $\mathcal{G}\ast \rho \equiv \int
\mathcal{G}(x-x') \rho(x',t) dx'$.

In summary, from model (\ref{geneq1}-\ref{releases}), the slow
signal-dynamics limit ($\bar{\delta}\ll \tau_R\ll
\tau_\phi,\tau_\rho$) leads to
\begin{align}\notag
\tau_\rho\partial_t \rho &= \mathcal{L}( \rho,\partial_x\rho) - \epsilon\rho\phi\, ,\\
\tau_\phi\partial_t \phi &= L( \phi,\partial_x\phi) + \bar R\rho  \, ,
\ \ \textrm{with} \ \ \bar R \equiv  \alpha \bar\delta I_0.
\label{sloweqs}
\end{align}
Fast signal dynamics $(\tau_\phi \ll
\bar{\delta} \ll \tau_R \ll \tau_\rho)$ gives
\begin{align}
\label{fasteq}
\tau_\rho\partial_t \rho &= \mathcal{L}( \rho,\partial_x\rho)
- \bar{\epsilon}\rho\, [\mathcal{G}\ast \rho] \, ,
\ \ \textrm{with} \ \ \bar{\epsilon}\equiv\epsilon \bar\delta\alpha \ .
\end{align}

Regardless of the choice of $\mathcal{L}(\rho,\partial_x\rho)$
and $L( \phi,\partial_x\phi)$, the fact that the two regimes
lead to different coarse-grained models suggests that their
spatial stability also differs. In fact, it can be shown that
pattern formation does not occur in the slow signal limit (Eq.
(\ref{sloweqs})) for a large class of operators (see \cite{SM}
for precise conditions on $\mathcal{L}$ and $L$). However, for
this same class in the fast limit, it is well known that Eq.
(\ref{fasteq}) can lead to spatial patterns when the signal
profile $\mathcal{G}$ is sufficiently
platykurtic~\cite{pigolotti2007,ricardo2013g}.

\textit{A particular example. ---}  We illustrate the above
developments with the following dynamics,
\begin{align}
\label{choice}
\mathcal{L}(\rho,\partial_x\rho) &= (D_\rho\partial_{xx} + r)\rho\, , \\ \notag
L(\phi,\partial_x\phi) &= D_\phi\partial_x(\phi^{\nu-1}\partial_x \phi) - [\gamma \phi^{\mu-1}]\phi\, ,
\end{align}
which models populations of organisms moving Brownianly with
diffusion coefficient $D_\rho$ and reproducing with growth rate
$r$. This choice is a fundamental building block for more
complex population dynamics models~\cite{murray2002}. For the
signal dynamics, Eq. (\ref{choice}) gives a generalized
nonlinear diffusion-decay process characterized by exponents
$\nu,\mu > 0$. It allows to consider the case where diffusion
and decay are sensitive to signal intensity in a negative
($\nu,\mu < 1$) or positive ($\nu,\mu > 1$) manner, unraveling
important channels through which environment structure (e.g.
propagation in porous media, leading to $\nu>1$, see
Ref.~\cite{hommel2018porosity}) and mediator inter-specific
biochemical interactions~\cite{murray2002,turchin,Cates2010,
okubo2013,foster2013,allee,Colombo2018} can affect signal
propagation dynamics.

\begin{figure}[t]
\includegraphics[width=\columnwidth]{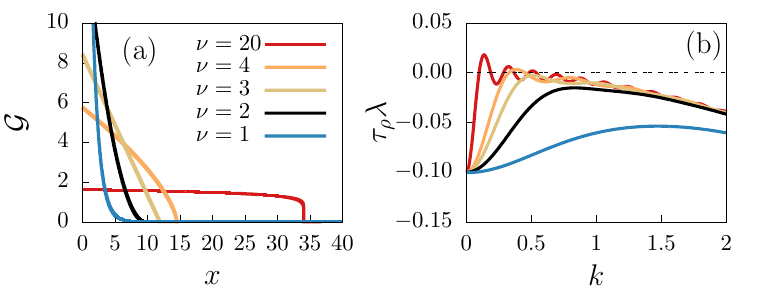}
\caption{(a) Profile $\mathcal{G}(x)$ from Eq. (\ref{kernel}) and
several $\nu$. (b) Corresponding
growth rates $\lambda(k)$ of perturbations to the homogeneous solution, as a function of
perturbation wavenumber $k$. Parameters are $D_\rho=0.01$, $\mu=1$, $\bar{\epsilon}=r=1$,
$D_\phi=\gamma=1$ and $I_0=10^2$.
}
\label{fig:stability}
\end{figure}


\begin{figure*}[t]
\includegraphics[width=2\columnwidth]{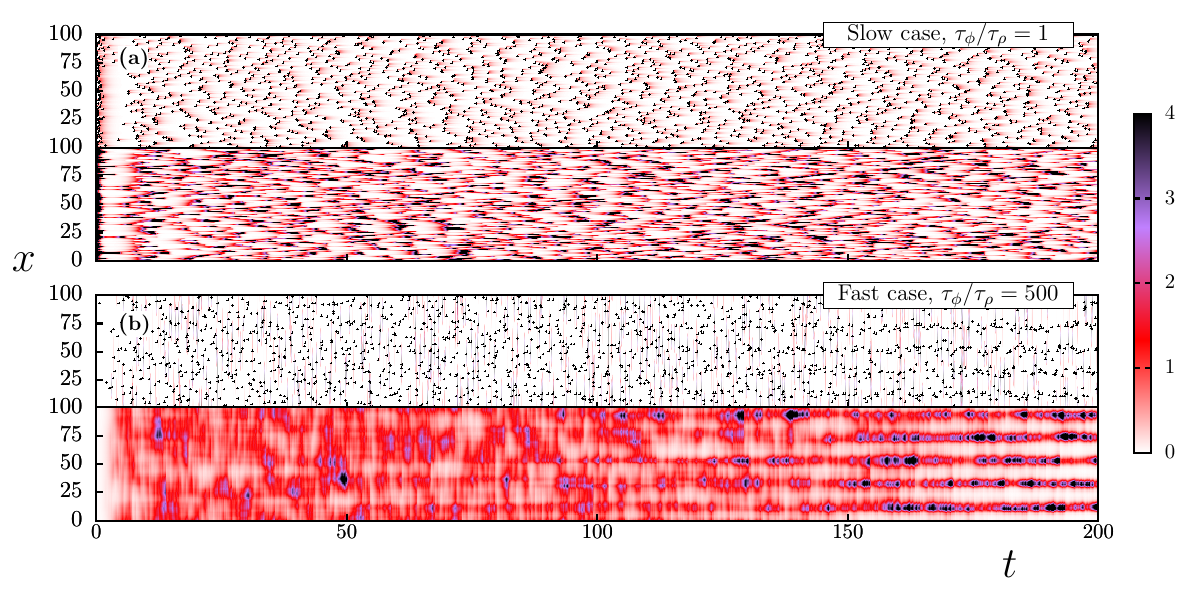}
\vspace{-0.5cm} \caption{ Temporal evolution of signal and
population fields for the slow (a) and fast (b) signal regimes.
The system is a line of length 100 with periodic boundary
conditions. Colors indicate field intensity $\phi(x,t)$ (upper
panels), and population density relative to the  (fast-signal limit) homogeneous
state, $\rho(x,t)/\rho_0$ (bottom panels). In the upper panels
crosses indicate release instants and positions
 (not all releases are captured by the
 finite resolution of the heatmap). Data from
numerical integration of Eqs.(\ref{geneq1}-\ref{releases}),
using an Euler scheme with $\delta t=10^{-6}$ and $\delta x =
1.0$. The dynamics is given by Eq.~(\ref{choice}) with
$D_\rho=10^{-2}$, $r=\tau_\rho=1$, $\alpha=10^2 $,
${\bar\delta}=10^{-2}$ and $\epsilon=10$. Signal dynamics is set by $\mu=1$ and $\nu=4$,
scaled by $\tau_\phi$, in such way that
$D_\phi/4=\gamma=1/\tau_\phi$ and $I_0=10^2/ \tau_\phi$. The
slow and fast regimes were obtained setting $\tau_\phi=1$ and
$\tau_\phi=1/500$, respectively. Mean interpulse interval is in
both cases $\tau_R = 1/(N(t)\alpha) \simeq 10^{-1}$ (for a
schematic close-up of the signal field in the fast-signal
regime see Fig.~\ref{fig:scheme}). }
\label{fig:poisson_patterns}
\end{figure*}

\emph{Linear stability analysis and pattern formation.---} The
pattern-forming stability conditions of model
(\ref{geneq1}-\ref{releases}) with the choice (\ref{choice})
can be obtained in the above studied timescale limits. For slow
signal dynamics (Eq. (\ref{sloweqs})) the
non-trivial homogeneous steady state is
$\rho_0=\gamma(r/\epsilon)^\mu/\bar R$, $\phi_0=r/\epsilon$.
Standard linear perturbation around this state identifies that
all perturbation growth rates are negative for
any value of parameters, implying the stability of the
homogeneous state. Hence, no pattern-forming instability can
arise.


For fast signal dynamics the model reduces to a single nonlocal
equation, Eq. (\ref{fasteq}), with integral kernel $\mathcal G$.
This is the solution of $L(\mathcal{G},\partial_x\mathcal{G})+
I_0\delta(x)=0$, an equation that can be solved exactly
\cite{Tsallis1996} for the particular choice (\ref{choice})
discussed here (additional details are in \cite{SM}) giving:
\begin{align}
\label{kernel}
\mathcal{G}(x) &= A
\left[ 1 - (1-q) \left| s x \right| \right]^{\frac{1}{1-q}}\, ,
\ A=\left[ \frac{I_0}{2 D_\phi}\sqrt{\frac{\mu+\nu}{2\kappa}} \right]^{\frac{2}{\mu+\nu}}\ ,
\end{align}
with $q = 1+(\mu-\nu)/2$, $s^2 = 2 \kappa A^{\mu-\nu}/(\mu +
\nu)$, and $\kappa=\gamma/D_\phi$. If $q < 1$ the support of
this solution is restricted to $|x|\leq 1/(1-q)$.

Fig.~\ref{fig:stability}a presents the different shapes of
$\mathcal{G}(x)$ as $\nu$ increases, while assuming linear
decay ($\mu=1$). The homogeneous steady solution is
$\rho_0=r/(\bar{\epsilon}\tilde{\mathcal{G}}(0))$, where
$\tilde{\mathcal{G}}(k)$ is the Fourier transform of $\mathcal
G$. Growth rates of periodic perturbations of wavenumber $k$ to
the homogeneous state are given by $\tau_\rho\lambda(k) = -
D_\rho k^2 - r\tilde{\mathcal{G}}(k)/\tilde{\mathcal{G}}(0)$
and are shown in Fig.~\ref{fig:stability}b. Pattern formation
requires that, for some $k$, $\tilde{\mathcal{G}}$ assumes a
sufficiently negative value, yielding $\lambda(k)>0$
\cite{pigolotti2007}. For the present case, this occurs if
toxin diffusion has a stronger sensitivity to concentration
when compared to the decay process, $\nu > \mu +2$. The
marginal case, $\nu = \mu+2$ ($\nu=3$ with $\mu=1$ in
Fig.~\ref{fig:stability}), corresponds to the triangular kernel
and the limit case $\nu\to\infty$ to the (most used) top hat
kernel, which is well-known to lead to pattern
formation~\cite{emilio1,Fuentes2003,Fuentes2004,Maciel21}.
Thus, in contrast to the slow signal-dynamics limit,  pattern
formation can occur under fast signal dynamics, showing the
importance of pulsed dynamics on the macroscopic behavior of
the system.

To support these analytical findings, we show in
Fig.~\ref{fig:poisson_patterns} direct numerical simulations
(see Supplemental Material for the numerical integration
scheme~\cite{SM}) of Eqs. (\ref{geneq1}-\ref{releases}) and
(\ref{choice}) (with $\mu=1, \nu=4$) for a slow (a) and a fast
(b) signal-dynamics regimes. This is done by keeping the
population timescale at $\tau_\rho = 1/r = 1$ for both plots,
and selecting the signal timescale corresponding to
$\tau_\phi=1\sim \tau_\rho$, and $\tau_\phi/ \tau_\rho = 1/500
\ll 1$, respectively. In agreement with the analytical results,
for the slow signal dynamics pattern formation does not occur
for any of the parameter values we have checked
(Fig.~\ref{fig:poisson_patterns}a). On the contrary, in the
fast limit patterns develop, since $\nu>\mu+2$.
Spatial population periodicity is seen to emerge at long times
in Fig. \ref{fig:poisson_patterns}b, and the spatial pattern
remains stable afterwards (see ~\cite{SM}). The wavelength of
the final pattern can be analytically estimated as
$2\pi/k^\star \simeq 16.5$, where $k^\star$ is the fastest
growing mode in Fig. \ref{fig:stability}b. This is roughly
close to the periodicity seen in
Fig.~\ref{fig:poisson_patterns}b (see also Fig. S2 in
Supplemental Material~\cite{SM}).

\textit{Final remarks and discussion.---} Our framework allowed
us to see how different fine-scale signal dynamics impact at a
coarser scale. It recovers standard
reaction-diffusion~\cite{murray2003,gierer1972} schemes in the
slow-signal limit and integrodifferential
schemes~\cite{pigolotti2007,o2020lamellar} in the fast-signal
limit, working as a bridge between the two
mostly used formalisms to describe interacting populations.

We crucially note that these two descriptions can lead to
different macroscopic outcomes. In this work, we focused on
showing that, for the same population and mediator dynamics,
a transition from slow to fast pulsed signals can effectively lead to spatially-extended
interference competition in such way that pattern formation
occurs~\cite{Fuentes2003,ricardo2014,pigolotti2007}.

Our findings are of relevance in situations, from chemistry to
ecology, in which interactions between the entities are
mediated by pulses that are short and fast compared to reaction
processes. More broadly, they stress crucial channels through
which environment and individual-level behavior can control
system spatial organization~\cite{fagan}. For example, our
approach can be extended to cases in which signals regulate
individual mobility~\cite{zinati2022stochastic,grimaPRL}, a
mediation that has already shown to be relevant for population
survival and spatial
patterns~\cite{lewis2007,giuggioli2011animal}. Developmental
programs can also explore these channels to engineer specific
morphologies~\cite{karig2018stochastic,grimaCurrent}. Further
extension aiming at concrete problems should include realistic
features such as: state-dependent signal
emissions~\cite{ojalvo2017,jordi2013} accounting for
individuals response to attacks; memory~\cite{potts2016memory};
persistence~\cite{giuggioli2011animal}; and multi-signal
mediation where signals establish a set of distinct biochemical
interactions~\cite{smith2003animal}.
\nocite{paulau2014selflocalized,oto2019,ruiz2020general}

\textit{Acknowledgements.---}
We acknowledge Ricardo Martinez-Garcia for a critical
reading of the manuscript.
This work has been supported by
the Severo Ochoa and Maria de Maeztu Program for Centers and
Units of Excellence in R\&D, grant MDM-2017-0711 funded by
MCIN/AEI/10.13039/501100011033.


\newcommand{\BE}{\begin{equation}}
\newcommand{\EE}{\end{equation}}
\newcommand{\BA}{\begin{eqnarray}}
\newcommand{\EA}{\end{eqnarray}}
\tolerance=1
\emergencystretch=\maxdimen
\hyphenpenalty=10000
\hbadness=10000
\renewcommand{\theequation}{S\arabic{equation}}
\renewcommand{\thefigure}{S\arabic{figure}}

\pagebreak
\setcounter{equation}{0}
\setcounter{figure}{0}
\setcounter{table}{0}
\setcounter{page}{1}
\onecolumngrid
\begin{center}
\large \textbf{Supplemental material: ``Pulsed interaction-signals as a route to pattern formation"}\\
Eduardo H.  Colombo, Crist\'obal
L\'opez and  Emilio
Hern\'andez-Garc\'{\i}a
\end{center}

\author{E. H.  Colombo}
\email{ecolombo@princeton.edu} \affiliation{Department of
Ecology \& Evolutionary Biology, Princeton University,
Princeton, NJ 08544, USA\looseness=-1} \affiliation{Department
of Ecology, Evolution, and Natural Resources, Rutgers
University, New Brunswick, NJ 08901, USA\looseness=-1}

\author{Crist\'obal
L\'opez}
\email{clopez@ifisc.uib-csic.es}

\author{Emilio
Hern\'andez-Garc\'{\i}a} \email{emilio@ifisc.uib-csic.es}
\affiliation{Instituto de F\'{\i}sica Interdisciplinar y
Sistemas Complejos (IFISC), CSIC-UIB, Campus Universitat de les
Illes Balears, 07122, Palma de Mallorca, Spain}

\section{Coarse-graining of the toxin field in the fast toxin-dynamics limit}

We demonstrated in the main text that the signal field in the
fast toxin-dynamics limit $\tau_\phi/\tau_\rho \rightarrow 0$
takes the approximate form
\BE
\phi(x,t) = \sum_i \mathcal{G}(x-x_i)\Pi_{\bar\delta}(t-t_i) \
,
\EE
where $\mathcal{G}(x)$ is the steady profile attained under a
single persistent pulse at $x=0$ and $\{t_i,x_i\}$ are the
starting times of the different toxin release events and their
locations. Here, we calculate the average or coarse-graining of
this field, $\left<\phi(x,t)\right>$, over small temporal and
spatial intervals:
\BE
\left<\phi(x,t)\right> \equiv \frac{1}{\Delta t \Delta x}
\int_{-\Delta t}^0\int_{-\frac{\Delta x}{2}}^{\frac{\Delta
x}{2}} \sum_i \mathcal{G}(x+x'-x_i)\Pi_{\bar\delta}(t+t'-t_i)
dt' dx' \ .
\EE
The temporal interval $\Delta t$ should satisfy $\Delta t \ll
\tau_\rho$ but we will also assume that it is larger than other
microscopic time scales: $\bar\delta, \tau_R \ll \Delta t$. At
difference with the slow toxin case, here the spatial
coarse-graining is not really needed, so that we eliminate it
from the expression by taking the limit $\Delta x\rightarrow
0$. The remaining temporal integral acts on the indicator
function $\Pi_{\bar\delta}$, selecting at any time $t$ only the
$m$ pulses that have occurred anywhere in the system during the
interval $[t-\Delta t,t]$. Thus we have:
\BE
\langle \phi(x,t) \rangle \approx
\bar\delta (\Delta t)^{-1} \sum_i^{m_R} \mathcal{G}(x-x_i) .
\EE
$m_R$ is the the total number of releases for the population which follows a Poisson distribution with mean $\alpha N(t) \Delta t$,
where $N(t)=\int \rho(x,t) dx$ is the total population size.
We have neglected the values of $t$ for which a pulse is only
partially contained in $\Delta t$. Because of the condition
$\bar\delta \ll \Delta t$ such time intervals are very small
and negligible at the population scale $\tau_\rho$. 

As in
the calculation for the slow signal case, the condition $\tau_R
\ll \Delta t $ implies a large value of $m_R$ so that, by the law
of large numbers, its fluctuations can be neglected. Also in
this case we can use $m_R^{-1}\sum_i \mathcal{G}(x-x_i) \approx
\int dx' \mathcal{G}(x-x') \textrm{pdf}_t(x')$, where
$\textrm{pdf}_t(x)=\rho(x,t)/N(t)$ is the probability density
of the locations $x_i$. Combining these results we arrive at
\BE
\left<\phi\right>\approx {\bar\delta}\alpha  \int
\mathcal{G}(x-x') \rho(x',t) dx' \equiv {\bar\delta}\alpha
[\mathcal{G}\ast\rho].
\EE

\section{Conditions for the absence of pattern formation in the slow signal-dynamics limit}

In this Section we establish conditions on the dynamics which
are sufficient to guarantee that pattern formation is absent in
the slow signal-dynamics limit. In this limit, Eqs. (5) of the
main text were found:
\BA
\partial_t\rho &=& \mathcal{L}(\rho,\partial_x \rho) - \epsilon\rho\phi \label{SMpopulation}\\
\partial_t\phi &=& L(\phi,\partial_x \phi) + \bar R\rho \ .
\label{SMtoxin}
\EA

By inspecting this structure we note that it is likely to not
produce patterns, since it has a linear coupling between the
equations~\cite{gierer1972}. To obtain the exact class of
operators for which we can guarantee that the homogeneous
solution of Eqs. (\ref{SMpopulation}-\ref{SMtoxin}) is stable,
we perform the following calculations.

To begin with, the steady and
homogeneous solution $(\rho_0>0,\phi_0>0)$ of (\ref{SMpopulation})-(\ref{SMtoxin}) is
\BE
\rho_0 = -\frac{L_0(\phi_0)}{\bar R} \ \ \ , \ \ \
\phi_0=-\frac{\bar R}{\epsilon}\frac{\mathcal{L}_0(\rho_0)}{L_0(\phi_0)}
\ ,
\EE
where we have defined $\mathcal{L}_0(\rho_0)\equiv
\mathcal{L}(\rho_0,0)$ and $L_0(\phi_0)\equiv L(\phi_0,0)$.
Note that positivity of $\rho_0$ requires $L_0(\phi_0)<0$.
 To check the stability properties, we linearize:
$\rho(x,t)=\rho_0+\delta\rho(x,t)$,
$\phi(x,t)=\phi_0+\delta\phi(x,t)$,
$\mathcal{L}=\mathcal{L}_0(\rho_0)+{\cal L}_1 \delta\rho(x,t) +
{\cal O}(\delta\rho)^2$, $L=L_0(\phi_0)+ L_1 \delta\phi(x,t) +
{\cal O}(\delta\phi)^2$.
Under Fourier transformation,
$\delta\rho(x,t)\rightarrow\delta\tilde\rho(k,t)$,
$\delta\phi(x,t)\rightarrow\delta\tilde\phi(k,t)$, ${\cal
L}_1\rightarrow { \cal \tilde L}_k$ and $L_1\rightarrow \tilde
L_k$, the linear stability of $(\rho_0,\phi_0)$ is guaranteed
if the following linear growth rates $\lambda_\pm(k)$ have
negative real parts $\forall k$:
\BE
\lambda_\pm(k) =\frac{1}{2} \left[  {\cal \tilde L}_k + \tilde
L_k -\epsilon\phi_0 \pm \sqrt{({\cal \tilde L}_k + \tilde L_k
-\epsilon\phi_0)^2 -4 ({\cal \tilde L}_k -\epsilon\phi_0)\tilde
L_k- 4 \bar R\rho_0} \right] \ .
\EE

We now impose additional restrictions on $\mathcal{L}$ and $L$
that would be sufficient to guarantee stability of
$(\rho_0,\phi_0)$. First, we assume that $L_k<0$. This is the
case if the dynamics of the toxin in the absence of release is
some diffusion-decay process. For the population dynamics
implemented in $\mathcal{L}$ we assume that the maximum of
${\cal \tilde L}_k$ is achieved at $k=0$, i.e. ${\cal \tilde
L}_k \le {\cal \tilde L}_{k=0}=\mathcal{L}_0(\rho_0)'$. This is
the typical case in which gradient terms in the population
dynamics are diffusion-like, but excludes models in which
higher order derivatives induce instabilities
\cite{paulau2014selflocalized,oto2019,ruiz2020general}.
Finally, we restrict to $\mathcal{L}_0(\rho_0)'\le
\mathcal{L}(\rho_0)/\rho_0$ ($= \epsilon\phi_0$). This is for
example the case if $\mathcal{L}_0(\rho_0)$ is linear, or the
well known logistic model
$\mathcal{L}_0(\rho_0)=a\rho_0-b\rho_0^2$ with $b \ge 0$. But
excludes the case $b<0$, which could lead to additional
homogeneous instabilities, as in
\cite{oto2019,ruiz2020general}. The stated conditions are
sufficient to guarantee that  ${\cal \hat L}_k + \hat L_k
-\epsilon \phi_0<0$, and then $\textrm{Re} \lambda_\pm(k)<0$
$\forall k$, and $(\rho_0,\phi_0)$ is stable so that no
pattern-forming instability occurs. In the main text we focus
on showing that, keeping the same dynamics for $\mathcal{L}$
and $L$, in the fast-signal limit the result for the linear
stability changes qualitatively and allows pattern formation to
occur for some classes of $\mathcal{L}$ and $L$.

\section{Stationary signal density field for a single release in the fast signal-dynamics limit}
Under the dynamics
\begin{equation}
\tau_\phi \partial_t \phi = L(\phi,\partial_x\phi) + \mathcal{R}_\rho(x,t) \, ,
\end{equation}
in the case of a single persistent and localized release, say
at $x=0$: $\mathcal{R}_\rho(x,t)=I_0\delta(x)$ if $t\in
[t_i,t_t+\bar\delta]$, and in the fast signal-dynamics limit
$\tau_\phi \to 0$, $\phi$ immediately achieves a stationary
profile $\mathcal{G}(x)$ which lasts while the pulse is present
($t\in [t_i,t_t+\bar\delta]$). Assuming that the signal
dynamics is ruled by
\begin{align}
\label{SMchoice}
L(\phi,\partial_x \phi) &= D_\phi\partial_x(\phi^{\nu-1}\partial_x \phi) - [\gamma \phi^{\mu-1}]\phi\, ,
\end{align}
the stationary profile $\mathcal{G}(x)$ satisfies
\begin{equation}\label{SMtoxEqpow}
D_\phi\partial_x (\mathcal{G}^{\nu-1} \partial_x \mathcal{G}) - \gamma \mathcal{G}^\mu = -I_0\delta(x)\, .
\end{equation}
This stationary solution can be found rewriting the above equation as
\begin{equation}\label{SMtoxEqpow2}
\partial_{xx} Z -  \nu \kappa Z^{\frac{\mu}{\nu}} =  \frac{-\nu I_0}{D_\phi} \delta(x)
\end{equation}
where $Z = \mathcal{G}^\nu$ and $\kappa=\frac{\gamma}{D_\phi}$.
Outside the release point we need that $\partial_{xx} Z =  \nu
\kappa Z^{\frac{\mu}{\nu}} $. The solution can be found
\cite{Tsallis1996} using as ansatz a generalization of the
exponential function, namely
\begin{align}\label{SMnlG}
 e_q(x) \equiv [ 1 - (1-q)| x|]^{\frac{1}{1-q}} \, ,
\end{align}
which is valid for $|x|\in [0,+\infty)$ for $q\geq 1$ and $|x|
\in \left[0,1/(1-q)\right]$ if $q<1$. This function recovers
the exponential function in the $q\to 1$ limit. Its derivative
is given by $ \frac{d e_{q}(sx)}{dx} = - s e_{q}^q $ , then,
consequently $ \frac{d^2 e_{q}(sx)}{dx^2} = s^2 q
e_{q}^{2q-1}\, . $ Hence, substituting $Z = A' e_{q'}(s'x)$ in
Eq.~(\ref{SMtoxEqpow2}) we find $s'^2 = (A')^{\mu/\nu
-1}\kappa\nu/q'$, $q'=\frac{\mu+\nu}{2\nu}$ and $A'= A^\nu$.
Using that $\mathcal{G} = Z^{1/\nu}$, we find that
\begin{align}
\label{SMkernel}
\mathcal{G}(x) &= A e_q(sx)\, ,\\ \notag
q = 1+\frac{\mu-\nu}{2},\, s^2 &= \frac{2 \kappa A^{\mu-\nu}}{(\mu + \nu)},\,
A = \left[\frac{I_0}{2D_\phi}\sqrt{\frac{\mu+\nu}{2\kappa}}\right]^{\frac{2}{\mu+\nu}} \ ,
\end{align}
where the value of the amplitude $A$ is found by considering the flux
constrain introduced by the point release, $\partial_x Z
|_{x=0}= -\nu I_0/(2 D_\phi)$. For $q<2$, the area under
the stationary profile is finite and it is given by $\tilde{ \mathcal{G}}(0)
\equiv \int dx \mathcal{G} (x)= \frac{2A}{(2-q)s}$.

\section{Numerical integration scheme}

In order to numerically integrate  Eqs. (1-3) of the main text,
we follow a standard forward Euler scheme complemented with the
generation of the stochastic state-dependent toxic release
$\mathcal{R_\rho}$. We discretize space in small cells of size
$\delta x$ and time in small intervals of duration $\delta t$
and define $\rho_{j,n}\equiv \rho(x=j\delta x, t= n \delta t)$,
and analogously with $\phi_{j,n}$. For the particular dynamics
given by Eq. (7) of the main text, the evolution after one
time-step $\delta t$ of the population and signal fields is
obtained as follows:
\begin{align}
\label{eq:numerical}
\rho_{j,n+1} &= \rho_{j,n} + [D_\rho(\rho_{j+1,n}+\rho_{j-1,n}-2\rho_{j,n})/(\delta x^2) +
r\rho_{j,n}-\epsilon \rho_{j,n}\phi_{j,n}] \delta t/\tau_\rho\, ,\\
\phi_{j,n+1} &= \phi_{j,n} + [D_\phi(\phi_{j+1,n}^\nu+\phi_{j-1,n}^\nu-2\phi_{j,n}^\nu)/(\nu\delta x^2)
- \gamma\phi_{j,n}^{\mu} + \mathcal{R}_{j,n}] \delta t/\tau_\phi\, .  \notag
\end{align}
We used that the nonlinear diffusion term
$\partial_x(\phi^{\nu-1}\partial_{x}\phi)$ can be written as
$\nu^{-1}\partial_{xx}\phi^\nu$ to help with numerical
instabilities. The stochastic variable $\mathcal{R}_{j,n}$ is
the discretized version the pulse release function
$\mathcal{R}_\rho(x,t)$ given by Eq.~(3) of the main text. It
is implemented as follows: Initially all $\mathcal{R}_{j,n}$
are set to zero. At each time step we check if a pulse will
occur somewhere in the system, with probability $\alpha N(t)
\delta t$, where $N(t)=\sum_k \rho_{k,n}\delta x$ is the total
population. If so, the needed pulse location $j$ is sampled
from its probability $\rho_{j,n}/\sum_k \rho_{k,n}$. Then, the
value $\mathcal{R}_{j,n}$ is set to $I_0/\delta x$ during
$\bar\delta/\delta t$ time steps, being reset to zero
afterwards. The denominator $\delta x$ arises from the
discretization of the spatial delta function. To ensure that
fast pulses are resolved by the numerical integration we
consider $\delta x=1.0$, smaller than signals' reach ($\sim
10$), and $\delta t=10^{-6}$, much smaller than the signal
duration time $\bar{\delta}=0.01$.

\section{Numerical steady solutions for fast signal dynamics}

\begin{figure}[h]
\begin{center}
\includegraphics[width=0.8\columnwidth]{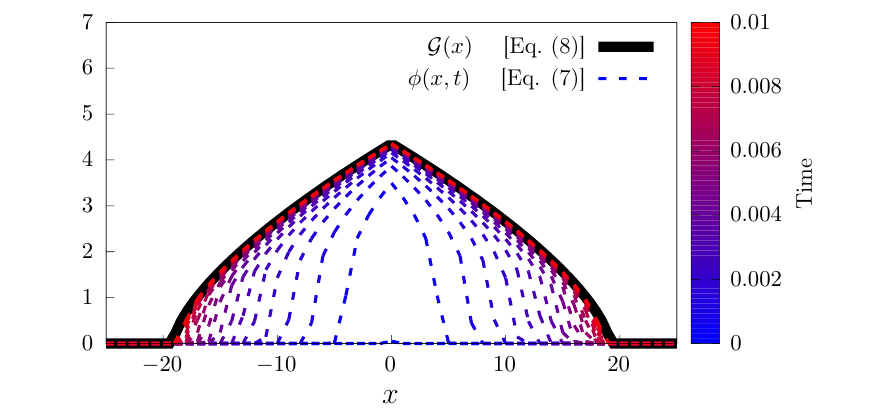}
\caption{Signal dynamics (dashed lines) from integration of Eq. (2) with (7) of the main text
starting from vanishing toxic signal,
but with signal release fixed to a delta function at the origin ($\mathcal{R}_\rho=I_0\delta(x)$).
Time is indicated by
color as shown in the colorbar. Parameters as in Fig. 3b
of the main text. After a short transient the profile correctly approaches the
stationary profile calculated in the fast signal-dynamics limit from the
integrodifferential description of Eq.~(8)
of the main text (solid black line).}
\label{fig:sm_numcheck}
\end{center}
\end{figure}

Both as a check of the numerical algorithm of the previous
Section, and to show the validity of the integrodifferential
approximation in Eq. (6) of the main text, obtained in the fast
signal-dynamics limit, we show in this Section two simulations
of the system (1-3) with (7) in a fast signal regime.

First, Fig.~\ref{fig:sm_numcheck} displays an integration of
Eq. (2) with (7) using the second equation of the numerical
algorithm (\ref{eq:numerical}) in a situation of fast signal
dynamics (parameters as in Fig. 3b of the main text) but with
the release fixed as a delta function at the origin
$\mathcal{R}_\rho(x,t)=I_0\delta(x)$
($\mathcal{R}_{j,n}=I_0/\delta x$ in the discretized version).
We see that the signal intensity profile correctly achieves in
a short time the analytical steady form given by Eq.~(8) of the
main text, which is adequate for this parameter regime.

\begin{figure}[h]
\begin{center}
\includegraphics[width=0.8\columnwidth]{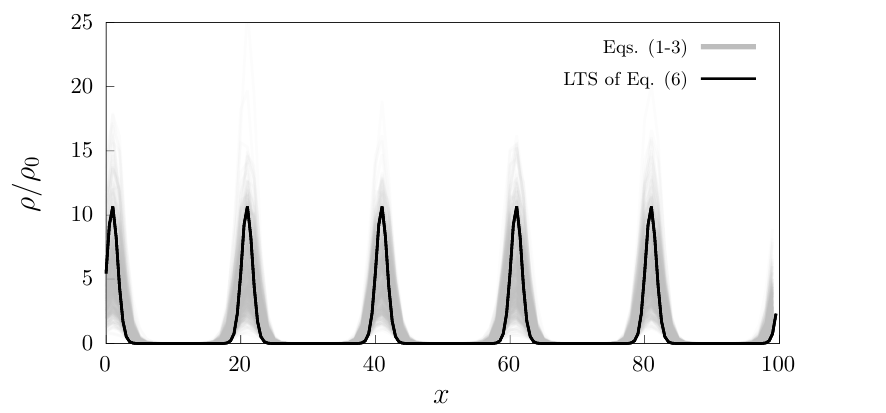}
\caption{Snapshots of the density profiles at the long-time regime are shown
as gray (95\% transparent) solid lines. Snapshots are from the same simulation as in
Fig. 3b of the main text, but for $200<t<1000$, being displayed at every unit time interval
(800 samples). The solid black line is the long-time state (LTS) of the integrodifferential Eq. (6),
appropriate for this fast signal-dynamics situation. This last solution
was spatially shifted so that the peaks of the two solutions match.}
\label{fig:sm_numcheck2}
\end{center}
\end{figure}

Second, in Fig.~\ref{fig:sm_numcheck2}, we show that the
pattern produced by the numerical integration of the stochastic
system (1-3) with (7) of the main text, in the fast-signal
regime (same simulation as in Fig.~3b of main text) matches the
stationary profile predicted by the integrodifferential
description Eq. (6) of the main text, obtained by numerical
simulation at long times, using random fluctuations around the
homogeneous solution as initial condition.

\end{document}